\documentclass[journal=jacsat,manuscript=article]{achemso}

\usepackage[version=3]{mhchem} % Formula subscripts using \ce{}
\usepackage{siunitx}
\usepackage{graphicx}
\graphicspath{{fig/}}
\usepackage{makecell}

\usepackage{geometry}
\usepackage{caption}
\usepackage{float}
\usepackage{natbib}
\usepackage{setspace}
\usepackage{xkeyval}
\usepackage{array}
\usepackage{booktabs}
\usepackage{threeparttable}
\usepackage{listings}
\usepackage{lmodern}
\usepackage{mathpazo}
\usepackage{microtype}
\usepackage{url}
\usepackage{hyperref}
\usepackage{fourier}
\hypersetup{breaklinks,colorlinks=true,linkcolor=blue,filecolor=blue,urlcolor=blue,citecolor=blue}

\usepackage{amsmath}
\usepackage{amssymb,amsfonts,latexsym,cancel}
\usepackage[compact]{titlesec}
\usepackage{xcolor}
\usepackage{calc}
\usepackage{titletoc}
\usepackage{lipsum}
\usepackage{multirow}
\usepackage{amstext}
\usepackage{tabularx}
\usepackage{adjustbox}
\usepackage[usetitle]{rsc}
\usepackage{epsfig,color}
\usepackage{placeins}
\usepackage{natmove}
\usepackage{titlesec}
\usepackage{tablefootnote}
\usepackage{graphicx,url}
\usepackage{bm}% bold math
\usepackage{physics}
\author{Rodrigo A. Mendes}
\affiliation{Quantum Theory Project, University of Florida, Gainesville, FL 32611, USA.}
\author{Peter R. Franke}
\affiliation{Quantum Theory Project, University of Florida, Gainesville, FL 32611, USA.}
\author{Ajith Perera}
\affiliation{Quantum Theory Project, University of Florida, Gainesville, FL 32611, USA.}
\author{Rodney J. Bartlett}
\email{bartlett@qtp.ufl.edu}
\affiliation{Quantum Theory Project, University of Florida, Gainesville, FL 32611, USA.}
\title[An \textsf{achemso} demo]
  {On the performance of QTP functionals applied to second-order response properties II: Dynamic polarizability and long-range C$_6$ coefficients}

\keywords{frequency-dependent polarizabilities, $C_6$ coefficients, COT, QTP}

\begin{document}

\begin{tocentry}

\includegraphics[width=1.0\linewidth]{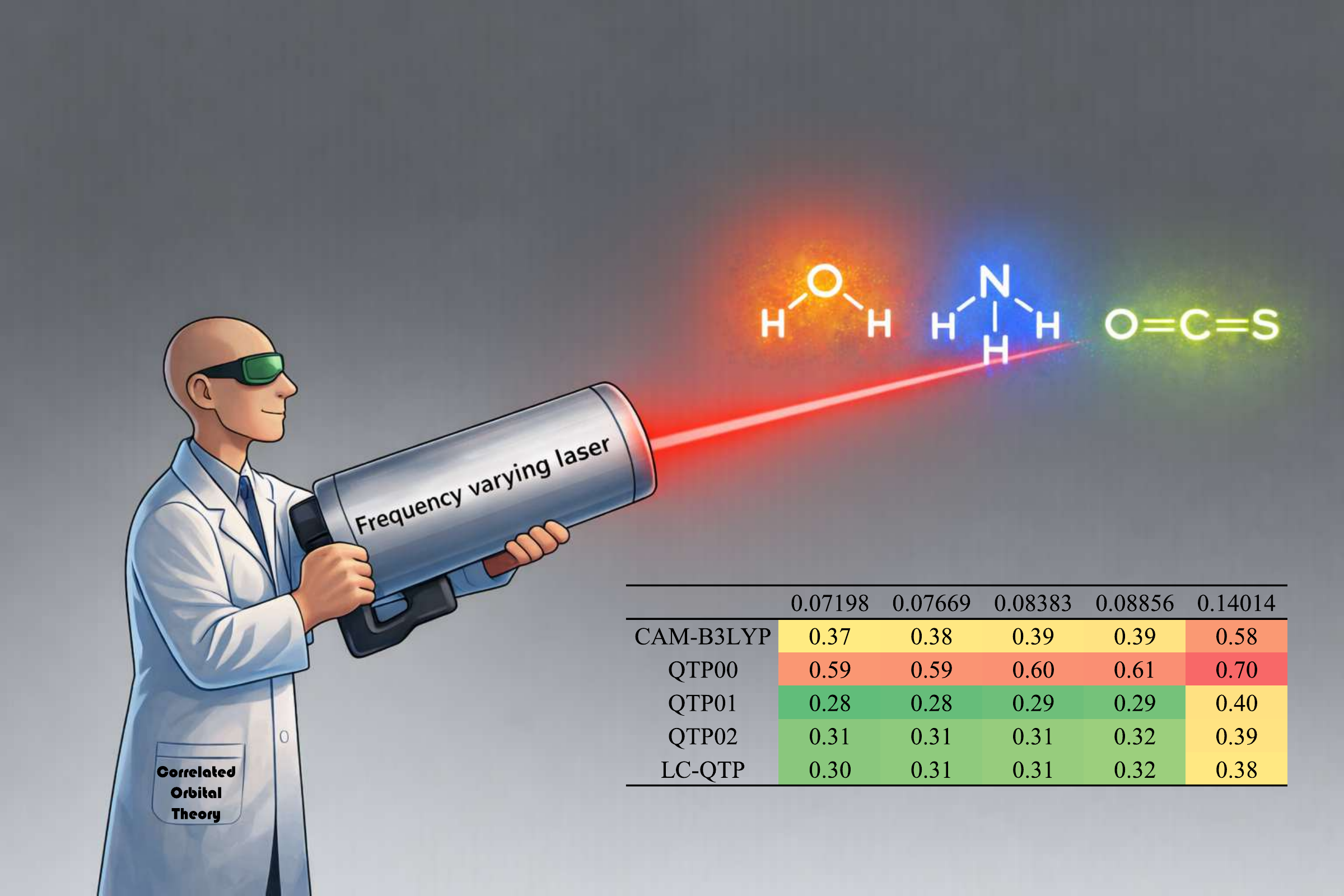}

\end{tocentry}

\begin{abstract}
This work is the second in the series “On the performance of QTP functionals applied to second-order response properties.” In the first paper (J. Chem. Phys. 162, 054105, 2025), we demonstrated the good performance of Quantum Theory Project functionals in predicting static perturbed second-order properties, such as static polarizabilities, nuclear magnetic resonance (NMR) spin-spin coupling constants, and NMR chemical shifts. In the present study, we focus on frequency-dependent properties, namely dynamic polarizabilities and $C_6$ dispersion coefficients. For completeness, a total of 25 exchange-correlation (XC) functionals were investigated. Dynamic polarizabilities were evaluated at five different perturbation wavelengths: 632.99 nm, 594.10 nm, 543.52 nm, 514.50 nm, and 325.13 nm. This property was also computed using HF and EOM-CCSD. In general, EOM-CCSD results are very close to those obtained with linear-response CC3, except at the highest frequency. Among Kohn-Sham calculations, TPSS0 and QTP01 showed the best overall performance for dynamic polarizabilities. We also assessed how well QTP functionals reproduce the pole structure of the CO molecule. For the $C_6$ dispersion coefficients, calculations were performed using the Casimir-Polder equation. The best overall performance was obtained with O3LYP; however, the first eleven ranked functionals show very similar accuracy. Within the QTP family, QTP01 and LC-QTP provide the best results for $C_6$ coefficients.
\end{abstract}

\section{Introduction}
The Correlated Orbital Theory (COT)\cite{BARTLETT_2009} can be viewed as a frequency-independent form of the Dyson equation from Electron Propagator (EP) theory, reformulated within the coupled-cluster (CC) framework.\cite{Meissner_1993} COT offers a coherent solution to the so-called ``Devil’s Triangle'' of Kohn–Sham Density Functional Theory (KS-DFT): the self-interaction error, the lack of integer discontinuity, and the incorrect one-particle spectra,\cite{Bartlett_2019,Mendes_2021} all which are different manifestations of the self-interaction error, but offer different route toward fixing the problem. According to COT, a one-particle theory should produce orbitals whose eigenvalues directly correspond to exact physical observables, including ionization potentials (IPs) for the occupied space and electron affinities (EAs) for some of the virtual orbitals (it depends on the number of stable anionic states in the specific chemical system). In this context, these orbitals are conceptually analogous to Dyson orbitals in EP theory, except they correspond to an infinite order sum provided by the frequency-independent self-energy from CC theory.

The functionals from the Quantum Theory Project (QTP) family, e.g., QTP00,\cite{verma2014increasing} QTP01,\cite{jin2016qtp} QTP02,\cite{haiduke2018qtp2} and LC-QTP\cite{haiduke2018qtp2}, are developed mapping KS-DFT onto COT. These functionals have proven successful in accurately describing a wide range of ground-state properties, such as ionization potentials (core and valence),\cite{Ranasinghe_2019,cQTP25} electron affinities,\cite{Pavlicek_2024} isotropic hyperfine coupling constants,\cite{Zack_hfcc_2022} and reaction barrier heights.\cite{haiduke2018qtp2} 

We showed in the previous paper of this series excellent performance from QTP functionals for static polarizabilities, NMR coupling constants, and chemical shifts for different isotopes, $^{13}$\ce{C}, $^{15}$\ce{N}, $^{17}$\ce{O}, $^{19}$\ce{F}, $^{31}$\ce{P}, and $^{33}$\ce{S}.\cite{Mendes_2025} For this, we investigated several functionals from the different rungs of Jacob’s ladder, always against EOM-CCSD results. Thus, LC-QTP ranked first in terms of the static polarizability. Regarding the total nuclear spin-spin coupling constants, QTP01 performed best (QTP02 and LC-QTP were second and third best). In the case of chemical shift analysis, TPSS0 surpassed the others for the majority of the chemical shifts. However, QTP00, QTP01, and QTP02 were among the best.

In addition, QTP functionals perform very well for various response properties, such as core excitation energies,\cite{Park_2022} vertical excitation energies (including Rydberg and charge-transfer states),\cite{Mendes_2021,mendes2021performance} two-photon absorption cross sections,\cite{Elayan_2024} and magnetizabilities.\cite{Lehtola_2021} Notably, QTP functionals have also demonstrated CC/EOM-CC quality in predicting both the fundamental and optical gaps of challenging extended systems, such as polyacene and trans-polyacetylene oligomers.\cite{Zack_2022,Bartaquim_2026}

In this second paper of our series on second-order properties, we examine how well QTP functionals describe frequency-dependent properties. For example, the dynamic polarizability can be expressed as the second derivative of the energy with respect to two electric fields, one static and one oscillating at frequency $\omega$, $-\frac{\partial^2 E}{\partial F_i(-\omega) \partial F_j(\omega)}$. This property measures how a molecule’s dipole moment responds to a time-varying electric field. Thus, the dynamic polarizability can be formally described by the sum-over-states (SOS) form,
\begin{equation}
\label{eq:SOS}
\alpha_{ij}(\omega) = \sum_{n} \left[ \frac{\matrixel{0}{\mu_i}{n} \matrixel{n}{\mu_j}{0}}{\omega_{n} - \omega} + \frac{\matrixel{0}{\mu_j}{n} \matrixel{n}{\mu_i}{0}}{\omega_{n} + \omega} \right]
\end{equation}
where $\mu_{i,j}$ is the Cartesian dipole moment component and $\omega_{n}$ is the excited state energy represented by excitations from $\ket{0}$ to an excited state $\ket{n}$, and is obtained from linear response TDDFT as,\cite{Casida_2012}
\begin{equation}\label{eq:tddft}
\left( \begin{matrix}
		A  &  B\\
		B^{\ast}  &  A^{\ast}\\
	\end{matrix}
	\right)  \left( \begin{matrix}
		X\\
		Y\\
	\end{matrix}
	\right) = \omega  \left( \begin{matrix}
		1  &  0\\
		0  &  -1\\
	\end{matrix}
	\right)  \left( \begin{matrix}
		X\\
		Y\\
	\end{matrix}
	\right).
\end{equation}
In this case, $X$ and $Y$ are particular eigenvectors having the $\omega$ eigenvalue for virtual-occupied and occupied-virtual elements of the density matrix. The matrices $\bm{A}$ and $\bm{B}$ have the following elements:
\begin{equation}
	\bm{A}_{ai,bj} = \delta_{ij} \delta_{ab} (\varepsilon_a - \varepsilon_i) + (ai|jb) + (ai|f_{xc}|jb),
\end{equation}

\begin{equation}
	\bm{B}_{ai,bj} = (ai|bj) + (ai|f_{xc}|bj),
\end{equation}
where the labels \textit{i,j} represent occupied orbitals, while \textit{a,b} denote virtual orbitals. $\varepsilon$ is the respective orbital eigenvalue. Furthermore, $f_{xc}$ is the ``kernel", \textit{i.e.}, the second derivative of the exchange-correlation functional (the adiabatic approximation is assumed),
\begin{equation}
f_{xc}=\frac{\delta^{2}E_{xc}}{ \delta  \rho  \left( 1 \right)  \delta  \rho  \left( 2 \right) }.
\end{equation}

In this work, we assess the performance of the QTP functionals in predicting the isotropic frequency-dependent polarizabilities under five distinct perturbations. We further examine their accuracy in describing the dispersion $C_6$ coefficients, which are directly related to the dynamic polarizability and provide information regarding the long-range interactions between chemical species. For completeness and a broader perspective, we also include a representative set of exchange-correlation (XC) functionals spanning the different rungs of Jacob’s ladder. Altogether, this results in a total of 25 XC functionals, in addition to the Hartree-Fock (HF) and coupled-cluster (CC) based post-HF methods.

\section{Computational methods}
The linear-response CC3\cite{cc3} calculations were performed using the Coupled-Cluster techniques for Computational Chemistry (CFOUR) software, version 1.2.\cite{cfour} Equation-of-Motion Coupled-Cluster with single and double excitations (EOM-CCSD)\cite{Stanton_1993} calculations were performed using the Advanced Concepts in Electronic Structure (ACES) II\cite{Perera_2020} package. No orbitals were frozen in any of the calculations. All DFT investigations were carried out using the PySCF package version 2.7.0\cite{pyscf1,pyscf2} combined with the LibXC exchange-correlation library.\cite{libxc} In this sense, the XC functionals of the following rungs of Jacob's ladder were investigated:\cite{Perdew_2001} Local Density Approximation (LDA), Generalized Gradient Approximation (GGA), meta-GGA (MGGA), global hybrids, and range-separated hybrids (RSH). For these calculations, we employed a numerical integration grid in PySCF with a grid level set to 6. This corresponds to Lebedev angular grids of 302 points for hydrogen, and 434 points for all heavier atoms, combined with Treutler-Ahlrichs radial grids (80 points for hydrogen, 120 points for second-row elements, and 125 points for third-row elements). The SCF convergence tolerance for the energy was set to \num{1.e-09}. All calculations were performed using the augmented triple-zeta Dunning basis set, aug-cc-pVTZ.\cite{Dunning_1989,Kendall_1992,Woon_1993} All the XC functionals used in this work are enumerated in Table{~}\ref{tb:xcs}.

\subsection{Isotropic Dynamic polarizabilities}
For the isotropic (average) dynamic polarizability calculations, we examined a set of 13 molecules: \ce{HF}, \ce{HCl}, \ce{H2O}, \ce{H2S}, \ce{NH3}, \ce{PH3}, \ce{CH4}, \ce{SiH4}, \ce{F2}, \ce{Cl2}, \ce{C2H4}, \ce{CO2}, and \ce{SO2}. The geometries were taken from the NIST Computational Chemistry Comparison and Benchmark Database and correspond to experimental coordinates.\cite{Russell_2018} The systems were perturbed at five different frequencies (a.u.): \num{0.071981} (632.99 nm), \num{0.076694} (594.10 nm), \num{0.083831} (543.52 nm), \num{0.088559} (514.50 nm), \num{0.140139} (325.13 nm). The isotropic values are obtained from,
\begin{equation}
    \Bar{\alpha} = \frac{1}{3} (\alpha_{xx} + \alpha_{yy} + \alpha_{zz})
\end{equation}

All isotropic dynamic polarizability values, along with the mean absolute deviation (MAD), mean signed deviation (MSD), maximum deviation (MAX), and the corresponding equations, are provided in the Supplementary Material.

\subsection{$C_6$ coefficients}
In the second part of this work, we investigated the long-range (attractive van der Waals) dispersion coefficients, $C_6$. The molecular set used for this analysis includes: \ce{C2H2}, \ce{C2H4}, \ce{CH3OH}, \ce{CH4}, \ce{Cl2}, \ce{CO2}, \ce{CO}, \ce{OCS}, \ce{CS2}, \ce{H2CO}, \ce{H2O}, \ce{H2}, \ce{H2S}, \ce{HBr}, \ce{HCl}, \ce{HF}, \ce{N2O}, \ce{N2}, \ce{NH3}, \ce{SiH4}, and \ce{SO2}. The geometries were retrieved from Cheng and Verstraelen.\cite{Cheng_2022}

To obtain the dipole polarizability of a system $S$ at imaginary frequencies, we write\cite{JIANG_2015}
\begin{equation}\label{eq:iSOS}
\alpha_S(i\nu) =
\sum_{n}
\frac{2 \omega_n \big|\matrixel{0}{\mu}{n}\big|^2}
{\omega_n^2 + \nu^2},
\end{equation}
where $\nu$ denotes an imaginary frequency. Note that in this case, the response function has no poles.

To compute the $C_6$ coefficients, we employed the Casimir-Polder formula,\cite{Casimir_1948} which in atomic units is given by
\begin{equation}\label{eq:cp}
C_6^{AB} = \frac{3}{\pi} \int_0^\infty \alpha_A (i\nu) \alpha_B (i\nu) d\nu.
\end{equation}

The integral in Eq.~\ref{eq:cp} was evaluated numerically using a Gauss-Legendre quadrature after mapping the semi-infinite interval [0, $\infty$) onto [-1, 1] via the transformation
\begin{equation}
\nu(x) = \nu_0 \frac{1 + x}{1 - x}, \quad x \in [-1,1].
\end{equation}
Under this change of variables, the integration measure becomes
\begin{equation}
d\nu = \frac{2\nu_0}{(1 - x)^2} dx.
\end{equation}

Thus, Eq.~\ref{eq:cp} can be rewritten as
\begin{equation}
C_6^{AB} = \frac{3}{\pi} \int_{-1}^1 \alpha_A\big(i\nu(x)\big) \alpha_B\big(i\nu(x)\big)
\frac{2\nu_0}{(1 - x)^2} dx,
\end{equation}
which is approximated as
\begin{equation}\label{eq:gl}
C_6^{AB} \approx \frac{3}{\pi} \sum_{i=1}^{n}
w_i \frac{2\nu_0}{(1 - x_i)^2} \alpha_A\big(i\nu_i\big) \alpha_B\big(i\nu_i\big),
\end{equation}
where $x_i$ and $w_i$ are the Gauss-Legendre quadrature points and weights, and $\nu_i = \nu(x_i)$

For this work, we used a 12-point Gauss-Legendre quadrature, which provided sufficient convergence, and adopted $\nu_0 = 0.3$, following the recommendations of Refs.~\citenum{Adamovic_2005,Fransson_2017}.
%%%%%%%%%%%%%%%%%%%%%%%%%%%%%%%%%%%%%%%%%%%%%%%%%%%%%%%%%%%%%%%%%%%%%%%%%%%%%%%%%%%

    \begin{table}[t!]
    \small
    \centering
    \caption{Exchange-correlation functionals used in this work (and HF method) and the respective percentage of non-local exchange.}
    \label{tb:xcs}
        \begin{threeparttable}
    \begin{tabular}{l l l l S l} 
        \toprule
	{XC}           & {Type}           && {\%$E_X^{HF}$}  & {$\omega_{RSH}$\tnote{a} (bohr$^{-1}$)}  &  {Ref.}	\\\midrule
    HF             & WFN              && 100             & {---}                     &  \citenum{hartree1928,fock1930,slater1930note}        \\
	SVWN5          & LDA              && 0	             & {---}                     &  \citenum{Slater1972,vwn}                             \\
	BLYP           & GGA	          && 0	             & {---}                     &  \citenum{becke1988,lee1988}                          \\
    BP86           & GGA	          && 0	             & {---}                     &  \citenum{becke1988,Perdew_1986}                      \\
	PBE            & GGA	          && 0	             & {---}                     &  \citenum{perdew1996generalized}                      \\
    XLYP           & GGA              && 0               & {---}                     &  \citenum{Xin2004,lee1988}                            \\
    OLYP           & GGA              && 0               & {---}                     &   \citenum{handy2001left,lee1988}                     \\        
	TPSS           & meta-GGA	      && 0	             & {---}                     &  \citenum{Tao2003}                                    \\
	M06-L          & meta-GGA	      && 0	             & {---}                     &  \citenum{Zhao2006}                                   \\   
    SCAN           & meta-GGA         && 0               & {---}                     &  \citenum{Sun_2015}                                   \\
    $r^2$SCAN      & meta-GGA         && 0               & {---}                     &  \citenum{Furness_2020}                                \\ 
    M11-L          & meta-GGA\tnote{b}     && 0	             & 0.25                      &  \citenum{Peverati_2012}                               \\     
	B3LYP          & GGA-hybrid       && 20              & {---}                     &  \citenum{becke1988,becke1993,lee1988,Stephens1994}   \\
 	PBE0           & GGA-hybrid       && 25	             & {---}                     &  \citenum{Adamo1999}                                  \\
	X3LYP          & GGA-hybrid       && 21.8            & {---}                     &  \citenum{Xin2004}                                    \\    
	O3LYP          & GGA-hybrid       && 11.61           & {---}                     &  \citenum{COHEN_2001}                                 \\    
	TPSSh          & meta-GGA-hybrid  && 10              & {---}                     &  \citenum{staroverov2003}                             \\ 
	TPSS0          & meta-GGA-hybrid  && 25              & {---}                     &  \citenum{Grimme_2005}                                \\ 
	CAM-B3LYP      & RSH	          && 19-65	         & 0.33                      &  \citenum{Yanai_2004}                                  \\
    LC-BLYP        & RSH              && 0-100           & 0.33                      &   \citenum{Iikura_2001}                                \\
	LC-PBEOP       & RSH	          && 0-100	         & 0.33                      &  \citenum{Iikura_2001}                                 \\
	LC-$\omega$PBE & RSH	          && 0-100	         & 0.40                      &  \citenum{Vydrov_2006}                                  \\
    QTP00          & RSH              && 54-99           & 0.29                      &  \citenum{verma2014increasing}                         \\
    QTP01          & RSH              && 23-100          & 0.31                      &  \citenum{jin2016qtp}                                  \\
    QTP02          & RSH              && 28-100          & 0.335                     &  \citenum{haiduke2018qtp2}                             \\
    LC-QTP         & RSH              && 0-100           & 0.475                     &  \citenum{haiduke2018qtp2}                             \\
	\bottomrule
    \end{tabular}
        \begin{tablenotes}
            \small
            \item[a] Range-separation parameter. In the literature, it is usually denoted by $\omega$ or $\mu$. Here, however, we use $\omega_{RSH}$ to avoid confusion with frequency and/or the dipole operator.
            \item[b] This functional is a special case of a range-separated functional, as it contains no HF exchange; instead, the separation occurs between different types of local exchange.
        \end{tablenotes}
    \end{threeparttable}    
\end{table}

\section{Results and discussion}
\subsection{Dynamic polarizabilities}
\begin{figure*}[t!]
    \centering
    \includegraphics[width=0.6\linewidth]{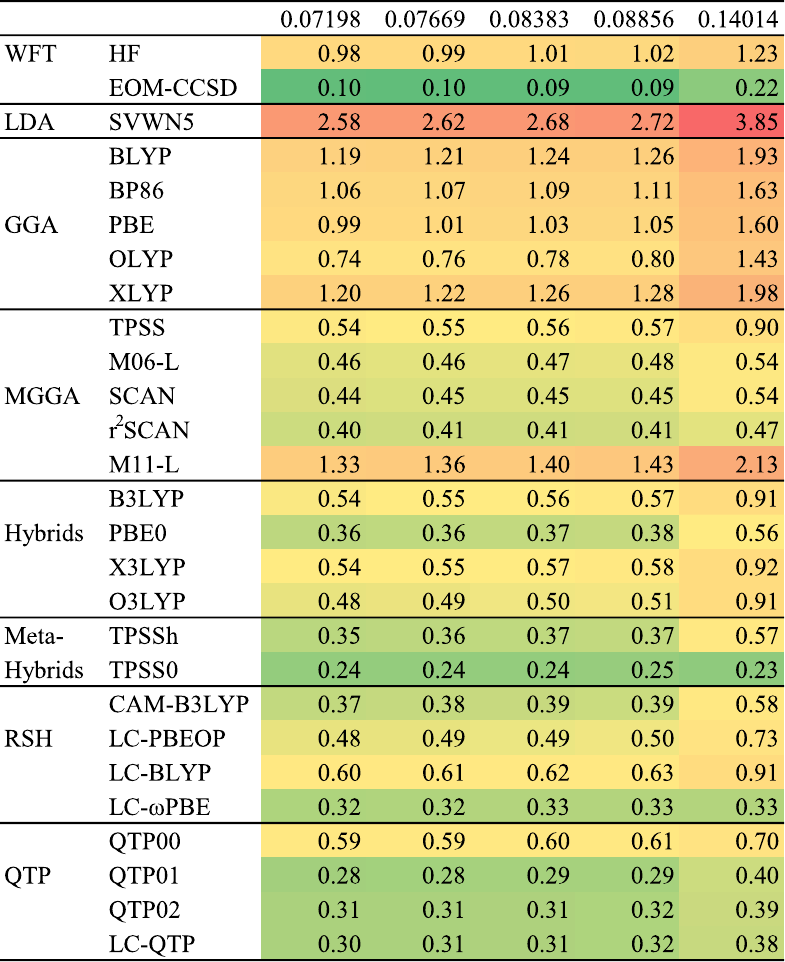}
    \caption{Heat map illustrating the mean absolute deviation (MAD) values in a.u., for the XC functionals compared to the LR-CC3/aug-cc-pVTZ level regarding dynamic polarizability.}
    \label{fig:dyna_all_freqs}
\end{figure*}

We evaluated the performance of the functionals and wavefunction methods by calculating the dynamic polarizabilities at five different frequencies. The mean absolute deviations (MADs) for all methods are presented in Figure~\ref{fig:dyna_all_freqs} as a heat map, with deviations calculated by comparison to linear response CC3 results. The best agreement is observed for EOM-CCSD, which deviates by only \SIrange{0.09}{0.10}{{a.u.}} from LR-CC3 for the first four frequencies. However, at the highest frequency, the deviation increases significantly, with a MAD of \SI{0.22}{{a.u.}} In general, the results reveal a consistent trend: all methods perform better at the four lowest frequencies in comparison to the highest frequency. This high-frequency error likely points to a poorer representation of the high-lying Rydberg states in the studied functionals, as well as some high valence states.

Among the HF, LDA, and GGA functionals, HF generally provides more accurate dynamic polarizabilities (\SIrange{0.98}{1.23}{{a.u.}}) than SVWN5 (MAD = \SIrange{2.58}{3.85}{{a.u.}}), BLYP (MAD = \SIrange{1.19}{1.93}{{a.u.}}), and XLYP (MAD = \SIrange{1.20}{1.98}{{a.u.}}). Also, none of the LDA or GGA functionals outperform HF at the highest frequency. Overall, within the LDA/GGA classes, OLYP shows the best performance, with MAD values ranging from \SIrange{0.74}{1.43}{{a.u}}. In a previous study,\cite{Mendes_2025} we observed a similar trend for the static polarizability, where HF produced more accurate results than the LDA and GGA functionals.

Much improved results are obtained as we move up Jacob’s ladder, as MGGAs generally exhibit significantly smaller MADs than GGAs. This suggests that the inclusion of second-order terms, such as the density Laplacian ($\nabla^2\rho$) or the kinetic-energy density ($\frac{1}{2} \sum_i^{occ}\mid \nabla\phi_i(\vec{r}) \mid^2$), is beneficial for second-order response properties under electric fields, consistent with both our previous observations for the static case and the findings of Hait and Head-Gordon.\cite{Mendes_2025,Hait_2018} As a result, the majority of the MGGA functionals show MAD values below \SI{1}{{a.u}}. Within this class, the best performer is r$^2$SCAN (MAD = \num{0.40}-\SI{0.47}{{a.u.}}), closely followed by SCAN (MAD = \num{0.44}-\SI{0.54}{{a.u.}}). Within this subset, the largest MAD values are obtained with M11-L (MAD = \num{1.33}-\SI{2.13}{{a.u.}}), which is the only poor performer among the MGGAs.

Among the GGA-hybrid functionals, PBE0 performs the best, with MAD values between \num{0.36}-\SI{0.56}{{a.u.}}, representing a clear improvement over all pure functionals from the lower rungs. The remaining hybrids, however, perform similarly to, or worse than, most MGGAs. Also, within the GGA-hybrid class, the performances are more uniform than in the previous ones. However, in this case, the worst performers are B3LYP (MAD = \num{0.54}-\SI{0.91}{{a.u.}}) and X3LYP (MAD = \num{0.54}-\SI{0.92}{{a.u.}}).

Once again, the inclusion of second-order ingredients on top of the density has a positive impact on the accuracy of frequency-dependent polarizabilities. TPSS0 delivers the best results, standing out even in the overall analysis (MAD = \num{0.24}-\SI{0.23}{{a.u.}}). The fact that TPSSh performs worse than TPSS0 (although it is among the best within this class and overall) suggests that the amount of HF exchange admixture (the only difference between them) is an important factor for this property: \SI{10}{\percent} in TPSSh versus \SI{25}{\percent} in TPSS0. It is worth noting, however, that this trend does not extend to the GGA-hybrid functionals, where a lower fraction of HF exchange does not necessarily correlate with lower performance.

None of the RSH functionals outperforms TPSS0. Focusing first on the ``non-QTP'' range-separated functionals, LC-$\omega$PBE provides the best MAD values (\num{0.32}-\num{0.33}) across all frequencies, while CAM-B3LYP performs comparably well at the lower frequencies (MAD = \num{0.37}-\SI{0.58}{{a.u.}}).

On the other hand, the best QTP functional, which is QTP01, performs as the second-best overall among the functionals, considering the first three frequencies the MAD are from \SIrange{0.28}{0.29}{{a.u.}}, and is slightly worse at the highest frequency, MAD = \SI{0.40}{{a.u.}} (the third-best overall for this frequency). Next, QTP02 (MAD = \num{0.31}-\SI{0.39}{{a.u.}}) and LC-QTP (MAD = \num{0.30}-\SI{0.38}{{a.u.}}) also perform very well among the RSH and overall. The worst performing QTP functional is QTP00, which has a MAD range of \SIrange{0.59}{0.70}{{a.u.}} for this property. 

The QTP functionals (except QTP00) generally perform very well for dynamic polarizabilities, as evidenced by the nearly constant and small MAD values for the first four frequencies. However, as observed for the majority of the other functionals, all QTP variants show a noticeable increase in error when the dynamic polarizability is evaluated at the highest frequency (\SI{0.14014}{{a.u.}}, corresponding to \SI{325.13}{nm}). Two functionals do not follow this trend, TPSS0 and LC-$\omega$PBE, which maintain consistently good performance across all frequencies.

\begin{figure*}[t!]
    \centering
    \includegraphics[width=0.9\linewidth]{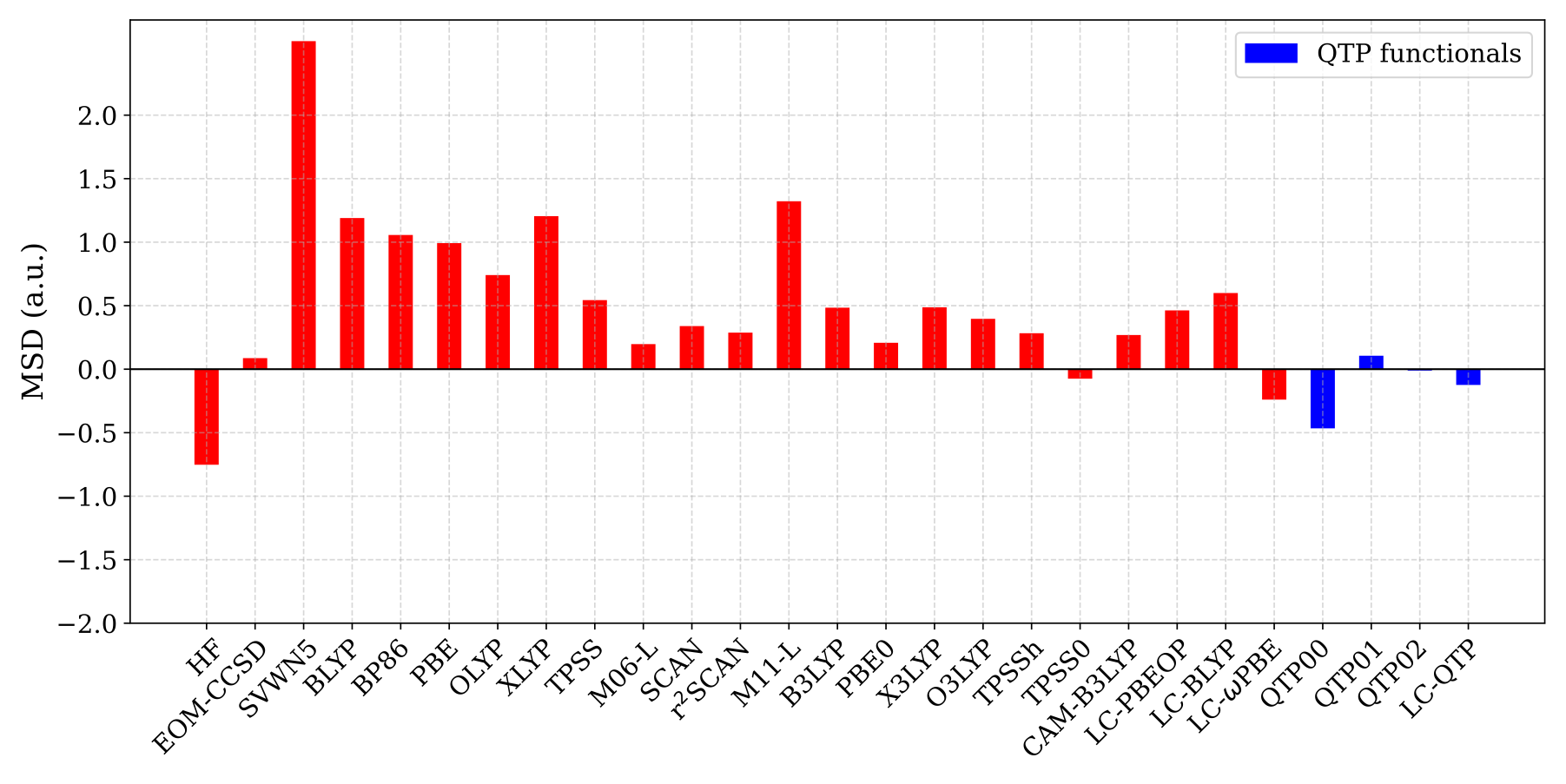}
    \caption{Mean signed deviation (MSD) values in a.u., for the XC functionals compared to the EOM-CCSD/aug-cc-pVTZ level for the dynamic polarizability.}
    \label{fig:dyna_msd}
\end{figure*}

Figure \ref{fig:dyna_msd} presents the mean signed deviation (MSD) values for the dynamic polarizability at the smallest frequency. MSD values for all other frequencies are provided in the Supplementary Materials. Overall, most functionals overestimate the dynamic polarizability. The largest deviation is observed for SVWN5, with an MSD of \SI{2.58}{{a.u}}. Among the pure GGA functionals, the MSDs are also relatively large, ranging from \SIrange{1.20}{0.74}{{a.u.}} for XLYP and OLYP, respectively. In contrast, the remaining functionals, i.e., meta-GGAs (except for M11-L), global hybrids, and RSHs, yield comparatively smaller MSD values. Furthermore, HF (MSD = \SI{-0.75}{{a.u.}}), TPSS0 (MSD = \SI{-0.07}{{a.u.}}), and LC-$\omega$PBE (MSD = \SI{-0.24}{{a.u.}}) all underestimate the dynamic polarizability. The QTP functionals also tend to underestimate, except for QTP01 (MSD = \SI{0.11}{{a.u.}}): QTP00 with an MSD of \SI{-0.47}{{a.u.}}, QTP02 with \SI{-0.01}{{a.u.}}, and LC-QTP with \SI{-0.12}{{a.u}}.

The maximum absolute deviations (MAX) reveal a clear pattern: across all functionals (and wavefunction methods), the largest errors are dominated by a small subset of molecules, namely \ce{SO2}, \ce{SiH4}, and \ce{C2H4}. All MAX values can be found in Tables{~}S2-S28 of the Supporting Information. For example, HF (MAX = \SIrange{1.90}{2.09}{{a.u.}}) and CCSD (\SIrange{0.31}{0.32}{{a.u.}}) show their largest deviations for \ce{SO2}. In contrast, LDA and GGA functionals (\SIrange{1.76}{7.17}{{a.u.}}) exhibit their largest errors for \ce{SiH4}. Most meta-GGAs (\SIrange{1.19}{5.77}{{a.u.}}) have their MAX associated with \ce{C2H4}. Among hybrids, two patterns emerge: PBE0 and O3LYP show their largest deviations for \ce{SiH4}, whereas B3LYP and X3LYP do so for \ce{C2H4}. MAX values within hybrids ranges from \SIrange{0.93}{1.98}{{a.u.}}. For meta-hybrids, the largest deviations occur for \ce{C2H4} (TPSSh and r$^2$SCAN) and \ce{SO2} (TPSS0), with MAX values spanning \SIrange{0.77}{1.37}{{a.u}}. Range-separated hybrids (including the QTP family) display more varied behavior, with their largest deviations distributed among all three key molecules. Thus, \ce{SO2} yields the largest errors for LC-$\omega$PBE, QTP00, QTP02, and LC-QTP (\SIrange{0.78}{1.52}{{a.u.}}), while CAM-B3LYP is dominated by \ce{C2H4}. The remaining functionals, i.e., LC-PBEOP, LC-BLYP, and QTP01, show their largest deviations for \ce{SiH4}, with MAX values ranging from \SIrange{0.89}{1.80}{{a.u}}.

Another important property that emerges from the analysis of the dynamic polarizability is the pole structure. Ideally, the poles of the response functions coincide with the corresponding excitation energies, and correspond to the residue at the poles in Eq.{~}\ref{eq:SOS}.

\begin{figure*}[t!]
    \centering
    \includegraphics[width=0.8\linewidth]{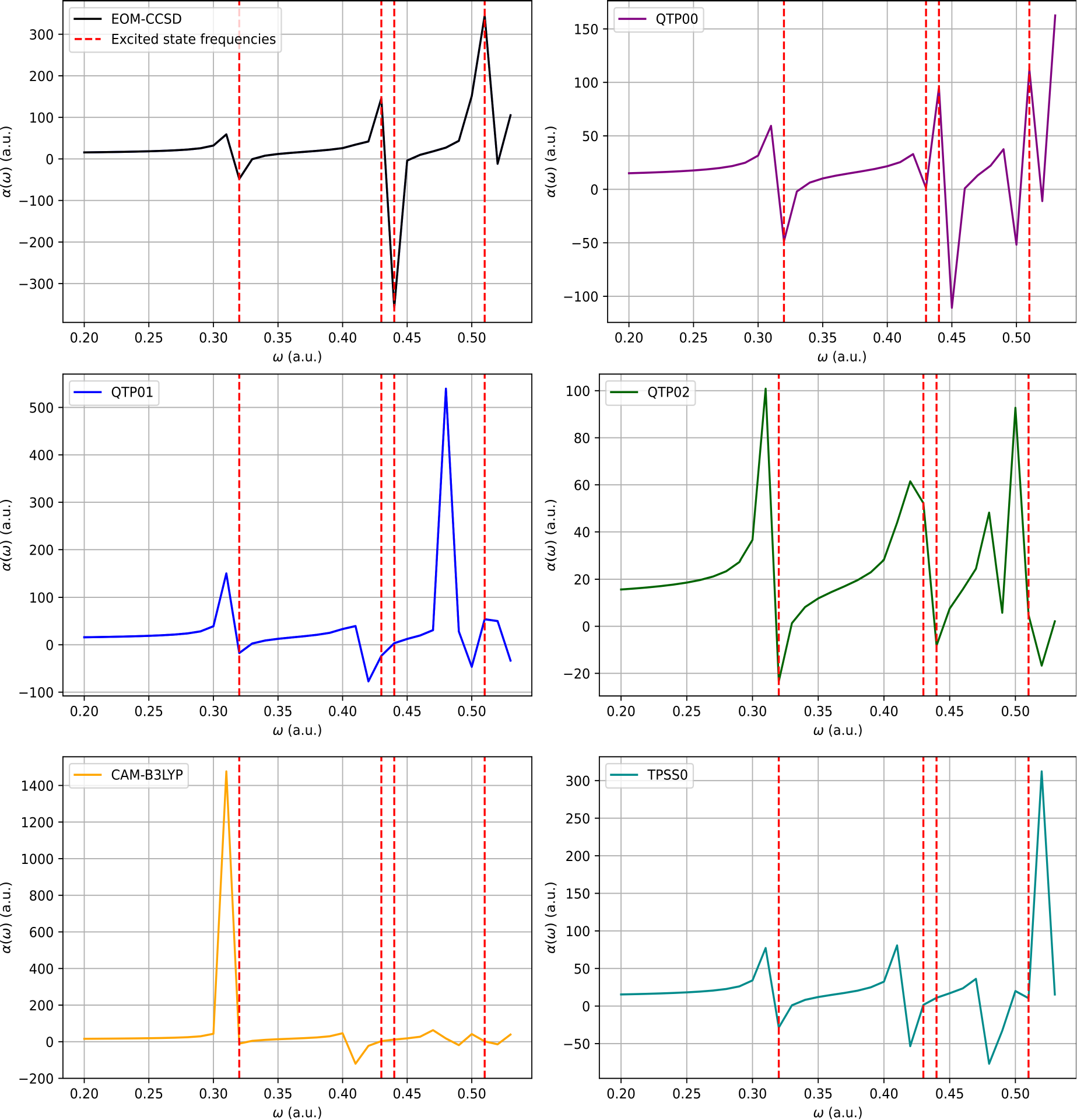}
    \caption{Isotropic dynamic polarizabilities for the \ce{CO} molecule ($r_e = $ \SI{1.1311}{\angstrom}),\cite{Ranasinghe_2019} computed at the range from $\omega_i = 0.20$ a.u. to $\omega_f = 0.53$ a.u. The basis set used was aug-cc-pVTZ.}
    \label{fig:poles}
\end{figure*}

Figure~\ref{fig:poles} illustrates the pole structure of the isotropic dynamic polarizability of the \ce{CO} molecule as obtained with EOM-CCSD (reference), TPSS0, CAM-B3LYP, and the QTP family of functionals. The figure displays polarizabilities in the range of \SIrange{0.20}{0.53}{{a.u.}} This range was chosen because the six lowest-lying bright ($f \ge$ 0.01) singlet excited states of the CO molecule fall within it, as calculated at the EOM-CCSD/aug-cc-pVTZ level: $^1\Pi$ \SI{8.60}{\electronvolt} (\SI{0.32}{{a.u.}} and $f=0.08$), $^1\Sigma^+$ \SI{11.76}{\electronvolt} (\SI{0.43}{{a.u.}} and $f=0.21$), $^1\Pi$ \SI{11.96}{\electronvolt} (\SI{0.44}{{a.u.}} and $f=0.06$), $^1\Sigma^+$ \SI{13.70}{\electronvolt} (\SI{0.50}{{a.u.}} and $f=0.20$), $^1\Pi$ \SI{13.66}{\electronvolt} (\SI{0.50}{{a.u.}} and $f=0.05$), and $^1\Pi$ \SI{13.88}{\electronvolt} (\SI{0.51}{{a.u.}} and $f = 0.10$). The vertical red dotted lines in Figure{~}\ref{fig:poles} indicate the positions of these excitation energies.

From the EOM-CCSD reference perspective, QTP00 shows the best comparison, while the CAM-B3LYP functional generally fails to reproduce both the positions and magnitudes of the dominant poles. In contrast, most exchange–correlation functionals perform reasonably well in reproducing the position of the first pole. The second pole, however, is systematically shifted by the DFT approximations. The QTP variants (which use CAM-B3LYP as a template) show closer qualitative agreement with the reference than CAM-B3LYP, exhibiting improved performance for both the first and second poles. Nevertheless, none of the functionals accurately reproduce the observed pole structure in the high-frequency region of the dynamic polarizability spectrum, where several spurious poles are also observed.

A qualitative analysis in the frequency range $\omega = 0.20-0.45$ shows that, for example, QTP00, QTP01, and TPSS0 exhibit the best performance, as they more accurately reproduce the overall pattern of the poles. In particular, these methods capture the nearly constant polarizability observed between \SIrange{0.32}{0.41}{{a.u.}}, a feature that is completely missed by the remaining functionals.

\subsection{Long-range dispersion \texorpdfstring{C$_6$}{C6} coefficients}
In this section, we evaluate the ability of the functionals to reproduce long-range dispersion interactions, quantified through the isotropic $C_6$ coefficients, for a set of 21 molecules. The $C_6$ coefficient represents an important term in the asymptotic expansion of the van der Waals dispersion energy and is a good test of how well a functional describes long-range electron correlation between fragments A and B. Our analysis compares the calculated isotropic $C_6$ values with experimental reference data. The performance of each functional is judged by the absolute and signed percentage errors with respect to the experiment. The explicit formulas used for these error metrics, along with the full set of calculated isotropic $C_6$ coefficients for all molecules and functionals considered, are provided in the Supplementary Material.

\begin{figure*}[t!]
    \centering
    \includegraphics[width=0.6\linewidth]{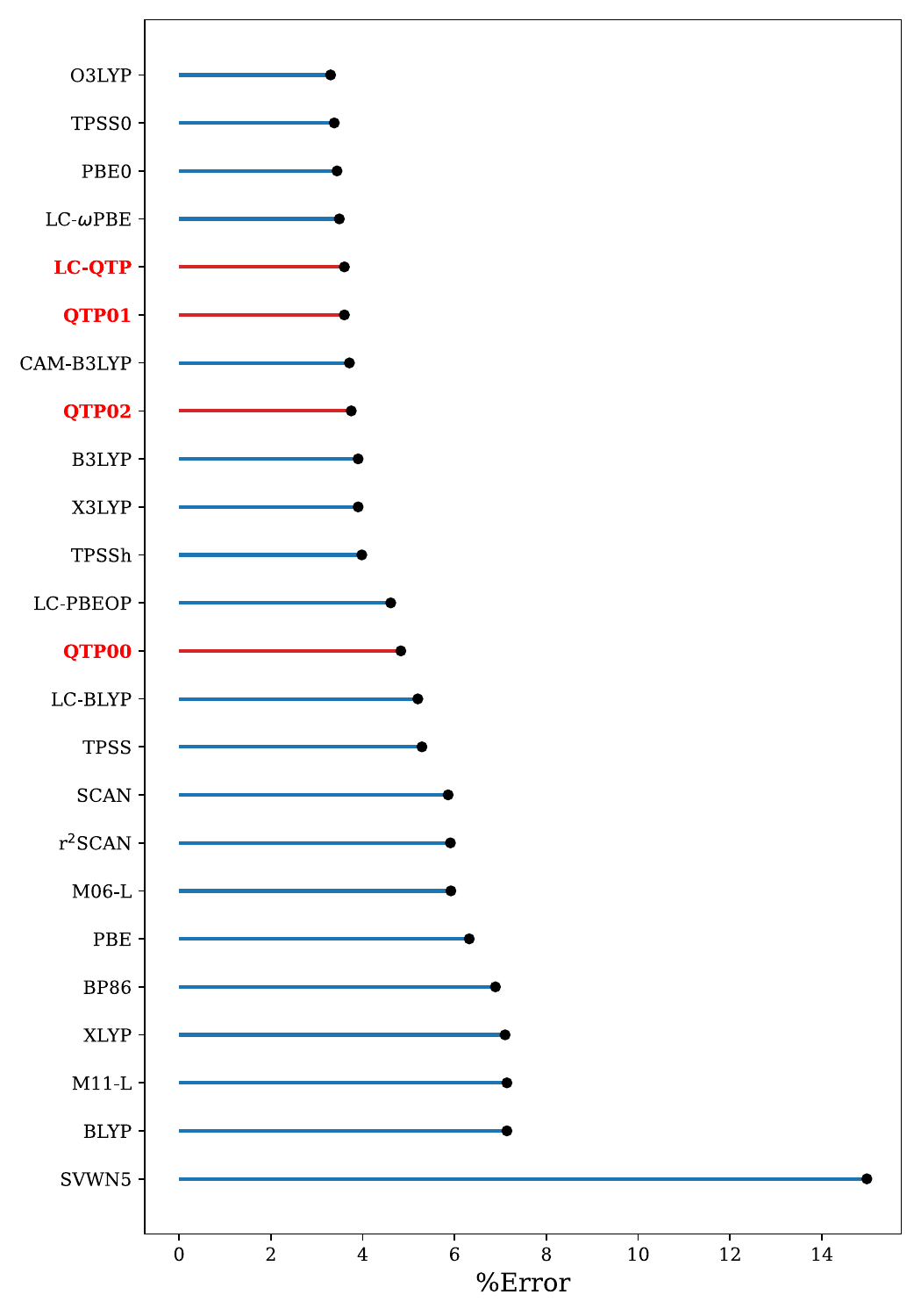}
    \caption{Percent error (\%Error) of isotropic $C_6$ coefficients obtained using different XC functionals w.r.t experimental values.}
    \label{fig:c6_error}
\end{figure*}

Figure~\ref{fig:c6_error} places the functionals in ranked order, with the best performer at the top and the worst at the bottom. Surprisingly, O3LYP emerges as the best-performing functional, deviating by only \SI{3.30}{\percent} from the experimental $C_6$ coefficients. The first eleven functionals in Figure~\ref{fig:c6_error} exhibit errors below \SI{4}{\percent}. The top three functionals are global hybrids and show very similar performance, with the second and third smallest errors of \SI{3.30}{\percent} (TPS00) and \SI{3.38}{\percent} (PBE0). Among the RSHs, LC-$\omega$PBE performs best, with \%Error = \SI{3.49}{\percent}. The best-performing QTP functionals are LC-QTP and QTP01, both yielding \%Error = \SI{3.60}{\percent}. These two members of the QTP family represent a slight improvement over CAM-B3LYP, which shows \%Error = \SI{3.71}{\percent}. QTP02 also ranks in the upper half of the list, with \%Error = \SI{3.75}{\percent}.

In the sequence, the worst-performing XC functional from the Quantum Theory Project family is QTP00, with \%Error = \SI{4.83}{\percent}. The poorest overall performance in Figure~\ref{fig:c6_error} is observed for SVWN5, which yields \%Error = \SI{14.98}{\percent}, representing a substantially larger deviation than that of all other functionals. Overall, the remaining functionals with errors exceeding \SI{4}{\percent} (other than SVWN5) exhibit \%Error values in the range \SIrange{4.61}{7.14}{\percent}.

\begin{figure*}[t!]
    \centering
    \includegraphics[width=0.9\linewidth]{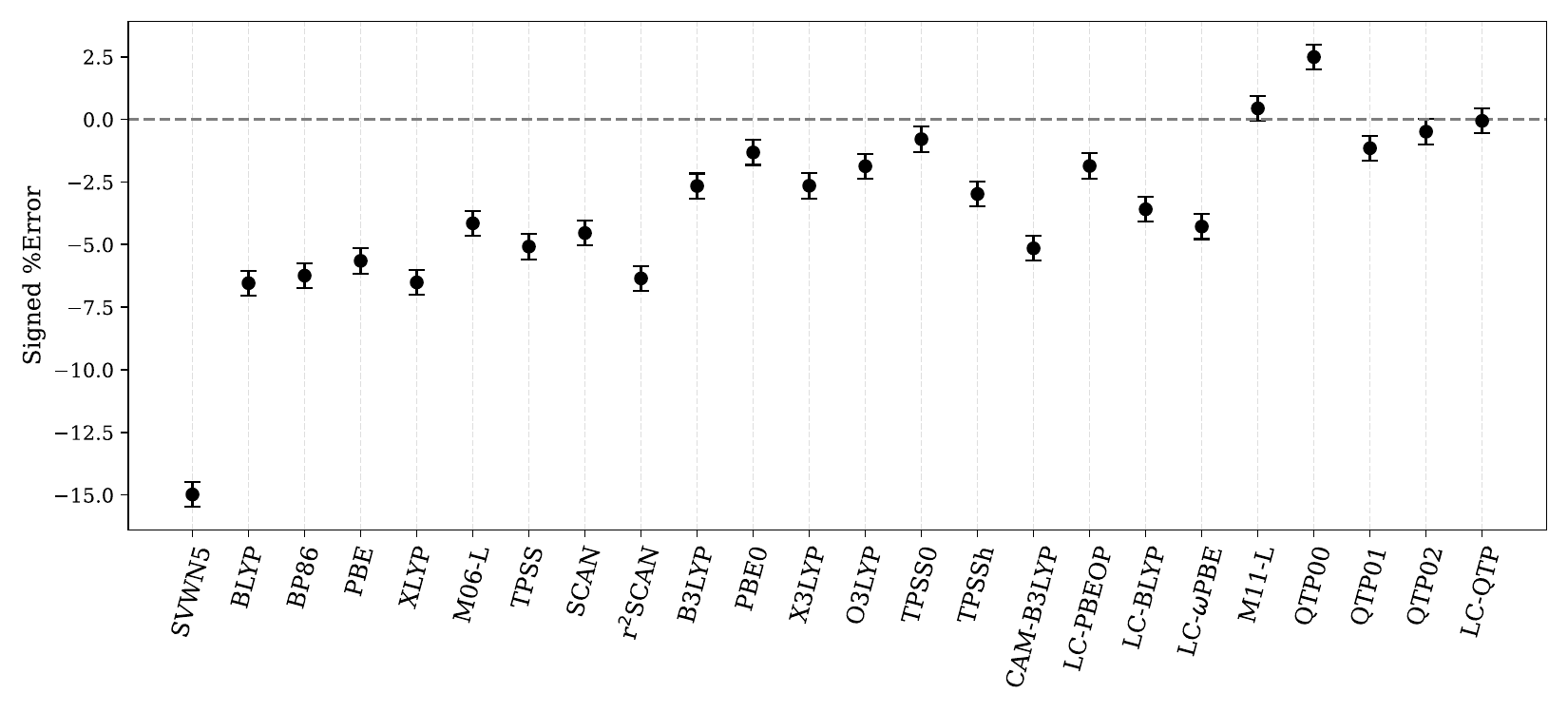}
    \caption{Signed percent error (\%Error) of isotropic $C_6$ coefficients obtained using different XC functionals w.r.t experimental values.}
    \label{fig:c6_sig}
\end{figure*}

Figure{~}\ref{fig:c6_sig} shows that most XC functionals underestimate the $C_6$ dispersion coefficients; exceptions are M11-L and QTP00. SVWN5 yields the poorest performance, with a signed percent error equal in magnitude to its absolute percent error, \SI{-14.98}{\percent}. GGA and meta-GGA functionals exhibit comparable signed errors, ranging from \SIrange{-6.54}{-4.15}{\percent} for BLYP and M06-L, respectively. In contrast, global hybrid functionals show signed errors closer to zero in comparison to the pure functionals, with values spanning \SIrange{-2.98}{-0.79}{\percent} for TPSSh and TPSS0, respectively. No clear systematic trend is observed among the RSH functionals. In this class, CAM-B3LYP shows the largest deviation, with a signed percent error of \SI{-5.15}{\percent}, whereas LC-QTP, with a signed percent error of \SI{-0.06}{\percent}, is the closest to zero among all XC functionals from Figure{~}\ref{fig:c6_sig}.

Finally, the Maximum percent errors for the $C_6$ coefficients (see Tables{~}S29-S34) reveal that two molecules dominate the largest deviations across the functionals: \ce{OCS} (11 functionals) and \ce{H2CO} (9 functionals). Two additional molecules appear less frequently, namely \ce{CS2} (1 functional) and \ce{H2} (3 functionals). Another noteworthy point is that, aside from a few outliers, the MAX values are relatively similar. The largest deviation is \SI{29.00}{\percent} (SVWN5), followed by \SI{17.19}{\percent} (M11-L). All remaining MAX values fall within a narrower range of \SIrange{10.12}{15.94}{\percent}, which corresponds to O3LYP and LC-BLYP, respectively.

\section{Conclusions}
We performed a careful and systematic investigation of the performance of 25 different XC functionals in predicting important second-order frequency varying properties in chemistry. These properties include dynamic polarizabilities, which were evaluated under perturbations at five different wavelengths, 632.99 nm, 594.10 nm, 543.52 nm, 514.50 nm, and 325.13 nm; as well as $C_6$ dispersion coefficients, which are obtained with the Casimir-Polder equation. For the isotropic dynamic polarizability, we also carried out calculations using three wavefunction methods: HF, EOM-CCSD, and LR-CC3. In this case, all comparisons were made against the LR-CC3 results obtained with Dunning's aug-cc-pVTZ basis set. For the isotropic $C_6$ coefficients, the performance of the XC functionals was assessed by comparison with experimental data.

In summary, for the isotropic dynamic polarizabilities, the excellent performance of EOM-CCSD at the first four frequencies indicates that triple corrections do not play an important role for the selected test set. However, the error approximately doubles when the performance of EOM-CCSD is evaluated at the highest frequency, i.e., 325 nm.

On the other hand, the best-performing XC functional is TPSS0, which exhibits consistent behavior across the applied frequencies. When comparing GGAs and MGGAs, the inclusion of second-order terms in the XC functional (i.e., $\nabla^2 \rho(r)$ or $\tau(r)$) appears to be beneficial for this type of molecular property. However, M11-L, which belongs to a range-separate MGGA class, does not seem to benefit from the inclusion of such terms. In general, the RSH functionals that perform best for dynamic polarizabilities are LC-$\omega$PBE and the QTP family, with the exception of QTP00. Notably, QTP01 is the second-best functional in this analysis, closely followed by LC-QTP. These two functionals, together with TPSS0, also showed the best performance in the first paper of this series (which investigated static polarizabilities, NMR coupling constants, and chemical shifts).\cite{Mendes_2025}

We also analyzed the pole structure associated with the first four low-lying excited states of \ce{CO} using EOM-CCSD, CAM-B3LYP, the QTP family, and the TPSS0 functional. In this analysis, all functionals except CAM-B3LYP closely follow the reference description for the first two poles; however, the pole at \SI{0.44}{{a.u.}} (highest frequency) is not accurately reproduced by any of the functionals considered.

Finally, for the isotropic $C_6$ dispersion coefficients, the first eleven functionals exhibit similar percent errors when compared with experimental results, with differences below \SI{4}{\percent}. In this case, the top-ranked XC functional is the hybrid O3LYP, closely followed by TPSS0, PBE0, LC-$\omega$PBE, LC-QTP, and QTP01.

\section*{Author contributions}
\textbf{Rodrigo A. Mendes}: 
Conceptualization (equal); 
Data curation   (lead); 
Formal analysis (lead); 
Investigation   (lead); 
Methodology     (lead); 
Writing original draft preparation (lead). 
\textbf{Peter R. Franke}: 
Investigation   (supporting);
Data curation   (supporting);
Writing– review \& editing (supporting). 
\textbf{Ajith Perera}:
Methodology (supporting);
Writing– review \& editing (supporting); 
\textbf{Rodney J. Bartlett}:
Conceptualization (equal);
Funding acquisition (lead);
Methodology   (supporting); 
Resources     (lead); 
Writing– review \& editing (supporting).

\section*{Conflicts of interest}
There are no conflicts to declare.

\section*{Data availability}
The data that support the findings of this study are available within the article and its supplementary material.

\begin{suppinfo}
The data employed for the Figures, complementary analyses, and additional technical details are reported within the Electronic Supplementary Material:
\end{suppinfo}

\begin{acknowledgement}
This work was supported by the Air Force Office of Scientific Research under AFOSR Award No. FA9550-23-1-0118. PRF was supported by the U.S. National Science Foundation, Division of Chemistry, under Award No. 2430408. We dedicate this work to John Stanton. John and RJB wrote a pioneering paper on using coupled-cluster (CC) theory for the dynamic polarizability (see ref.{~}\citenum{Stanton_5178_1993}). And, once again, we address that question with COT that derives from CC theory, and takes us into the domain of an exact correlated orbital theory.
\end{acknowledgement}

\bibliography{refs}

\end{document}